\documentclass[11pt,nofootinbib,floatfix]{revtex4}
\usepackage{mathrsfs}
\usepackage{graphicx}
\usepackage{amsmath}
\usepackage{amsfonts}
\usepackage{amssymb}
\usepackage{color}

\usepackage{epsfig}
\usepackage{CJK}
\usepackage{eepic}
\usepackage{bbm}
\usepackage{dcolumn}
\usepackage{bm}
\usepackage{ulem}
\usepackage{slashbox}
\usepackage{multirow}

\def\Lie{\pounds}

\newcommand{\omits}[1]{}

\def\bc{\begin{center}}
\def\nno{\nonumber}
\def\ec{\end{center}}
\def\be{\begin{eqnarray}}
\def\ee{\end{eqnarray}}

\def\vl{\ell_Q}
\providecommand{\norm}[1]{\lVert#1\rVert}

\def\de{\delta_e}
\def\tt{\mathbf{t}}
\def\ss{\mathbf{s}}
\def\vv{\mathbf{v}}

\definecolor{dyellow}{rgb}{1.,0.8,.0}
\definecolor{myblue}{rgb}{.1,.1,.7}
\definecolor{dcyan}{rgb}{.0,.6,.6}
\definecolor{cyan}{rgb}{0.4,1.0,1.0}
\definecolor{dmagenta}{rgb}{0.6,0.0,0.6}
\definecolor{brown}{rgb}{0.6,0.2,0.}
\definecolor{darkblue}{rgb}{.0,.0,0.5}
\definecolor{darkred}{rgb}{0.75,0.0,0.0}
\definecolor{orange}{rgb}{1.,.6,.0}
\definecolor{dorange}{rgb}{0.8,.4,.0}
\definecolor{green}{rgb}{0.0,1.0,0.0}
\definecolor{darkgreen}{rgb}{0.0,0.6,0.0}
\definecolor{purple}{rgb}{.4,.0,.4}
\definecolor{lightgrey}{rgb}{0.7, 0.7, 0.7}
\definecolor{grey}{rgb}{0.4, 0.4, 0.4}

\begin{document}


\title{Note on acoustic black holes from black D3-brane}

\author{Jia-Rui Sun$^{1}$} \email{sunjiarui@mail.sysu.edu.cn}
\author{Chao Yu$^{1}$} \email{yuch8@mail2.sysu.edu.cn}

\affiliation{${}^1$School of Physics and Astronomy, Sun Yat-Sen University, Guangzhou 510275, China}



\begin{abstract}
  In this paper, we study the acoustic black hole emerged from the nonextremal black D3-brane, based on the holographic approaches in constructing the acoustic black hole in asymptotically Anti de-Sitter spacetime (AAdS) and the effective hydrodynamic description of the nonextremal black D3-brane. We show that the holographic dual description of the acoustic black hole appeared on the timelike cutoff surface in the nonextremal black D3-brane also exist. The duality includes the dynamical connection between the acoustic black hole and the bulk gravity, a universal equation relating the Hawking-like temperature and the Hawking temperature, and a phonon/scalar channel quasinormal mode correspondence.

\end{abstract}


\maketitle
\newpage

\section{Introduction}\label{sec:intro}
The analogy gravity is an interesting phenomena that usually appeared in nongravitational systems such as the normal nonrelativistic fluid~\cite{Unruh:1980cg,Visser:1997ux}, the superfluid~\cite{Volovik:2000ua,Volovik:2002ci}, the Bose-Einstein condensate (BEC)~\cite{Garay:1999sk,Barcelo:2001ca}, the optical fibers~\cite{Belgiorno:2010zz,Belgiorno:2010iz} and the relativistic fluid~\cite{Bilic:1999sq,Visser:2010xv,Ge:2010wx,Ge:2010eu} etc, see the review article~\cite{Barcelo:2005fc} for more examples. The typical features are the emergence of the curved geometry and the Hawking-like radiation, which resemble the curved spacetime and quantum effects in it. These properties are interesting in that they suggest possibilities to build analogous gravitational model in the laboratory to mimic the real gravity, say, a black hole or the universe. In addition, they may also indicate an emergent picture of the gravity~\cite{Hu:2005ub,Gu:2009jh,Xu:2010eg,Sindoni:2011ej}. An obvious problem is that, although the analogy gravity exhibit many features like the real gravity, their underlying dynamics looks rather different with that of the gravity, i.e., the Einstein equation. For example, the dynamical equation which governs the fluid is the Navier-Stokes equation, the equation of motion governs the dynamics of the superfluid is the Gross-Pitaevskii equation and the fibre optics is characterized by the Maxwell equations. Therefore, the analogy gravity is mainly regarded as the model that analogous to the real gravitational systems since its discovery. Nevertheless, they still provide possibilities to test important phenomena of the real gravitational systems in the laboratory. Actually, experimental construction of the analogy gravity are realized in some cases recently, e.g., the measuring of the stimulated ``Hawking radiation'' in acoustic white hole in fluid~\cite{Weinfurtner:2010nu} and the observation of self-amplifying ``Hawking radiation'' in analogous black hole in BEC~\cite{Steinhauer:2014dra}. Both experiments claimed the detection of the Hawking radiation. However, as mentioned above, without further dynamical connections between the analogy gravity and the real gravity, they can only be called the Hawking-like radiation. An interesting question is, to what degree can the analogy gravity reflects the real gravity? Answering this question is important because it determines whether the analogy gravity is merely an analogy or not. In a recent paper~\cite{Ge:2015uaa}, the authors found that one of the analogous gravitational system--the acoustic black hole in normal fluid, can indeed be mapped to the real gravitational object--a black hole in asymptotically Anti-de Sitter (AAdS) spacetime from a holographic perspective. More specifically, the dynamical connection between the acoustic black hole and the AdS black hole, the connection between the Hawking-like temperature and the Hawking temperature, as well as the duality between the phonon propagating in the acoustic black hole and the scalar channel quasinormal mode propagating in the AdS black hole have been revealed for the first time. Based on this study, the acoustic black hole formed in fluid is no longer just an analogy, it really corresponds to a real black hole in a higher dimensional AdS spacetime. Furthermore, This finding gave strong support to the experimental efforts in making analogous gravitational systems in the laboratory. Other attempts in finding the relation between the analogy gravity and the real gravitational systems can be found in~\cite{Das:2010mk,Hossenfelder:2015pza,Hossenfelder:2014gwa,Dey:2016khw} where the relationship between the acoustic metric and the black hole solution are analyzed in some examples.

As we have mentioned above, there are many other analogous gravitational systems apart from the class studied in~\cite{Ge:2015uaa}. In the present paper, we would like to study another interesting system, which is the acoustic black hole emerged from the black D3-brane in type IIB string theory. Unlike the AdS black hole, the black D3-brane (non-extremal case) is in the asymptotically flat spacetime in which the aspects of holographic duality is not so clear as the well-established AdS/CFT correspondence. However, it is shown that the non-extremal black D3-brane also behaves like a hydrodynamic system at the specific timelike cutoff surfaces in the low frequency and long wavelength limits~\cite{Emparan:2013ila}. Therefore, it is expected that the acoustic black hole appeared on the timelike cutoff surface has a gravity dual description. By applying the approaches developed in~\cite{Ge:2015uaa} we show that when an acoustic black hole forms in the fluid at the finite cutoff timelike surface in the black D3-brane, it corresponds to a real gravitational system in the asymptotically flat spacetime which is a perturbed black D3-brane, as expected. The mapping is built upon the matching between the dynamical equations governing the acoustic black hole and those determining the quasinormal mode in the bulk gravity.

The organization of the paper is as follows. In section~\ref{sec:D3 brane} we briefly review basic properties of the black D3-brane and its effective hydrodynamic description. In section~\ref{sec:acoustic bh} we analyze the properties of the acoustic black hole which comes from the hydrodynamic system at the finite timelike cutoff surface and study its dual gravitational counterpart in the asymptotically flat spacetime. In section~\ref{sec:bulk boundary} we show that the Hawking-like temperature of the acoustic black hole is connected to the Hawking temperature of the bulk black D3-brane. Furthermore, we obtain the field/operator duality relation between the phonon propagating in the acoustic black hole and the sound channel of the quasinormal mode in the bulk perturbed black D3-brane. Finally, we conclude in section~\ref{sec:conclusion}.

\section{The black D3-brane and its hydrodynamic description}\label{sec:D3 brane}
\subsection{The black D3-brane solution}
The black D3-brane is a black hole solution of the type IIB supergravity in 10-dimensional flat spacetime which is the low energy effective theory of the type IIB string theory, it has the form as~\cite{Horowitz:1991cd}:
\be
    ds^2 &=& - \Delta_+ \Delta_-^{-\frac{1}{2}} dt^2 + \Delta_-^{\frac{1}{2}} dy_i^2 + \frac{dr^2}{\Delta_+ \Delta_-} + r^2 d\Omega_5^2 \nno\\
         &\equiv& ds_5^2 + r^2 d\Omega_5^2,\nno\\
    F_{[5]}&=&\kappa_5^2 Q(1+*){\rm Vol}(S^5),
\ee
with
\be
    \Delta_\pm = 1-\frac{r_\pm^4}{r^4} \quad {\rm and}\quad i=1,2,3,
\ee
in which $F_{[5]}$ is the 5-form field strength associated with the Ramond-Ramond (R-R) 4-form field, $*$ is the Hodge star, $r_\pm$ are respectively the inner- and outer-horizon radius of the black D3-brane and they are related to the black brane R-R charge as
\be
    Q = \frac{2\Omega_5}{8\pi G_{10}} r_+^2 r_-^2 = \frac{2}{\kappa_5^2} r_+^2 r_-^2
\ee
where $G_{10}$ is the 10-dimensional Newtonian coupling constant and $\kappa_5^2\equiv 8\pi G_{10}/\Omega_5$ its reduced 5-dimensional counterpart.

\subsection{The boundary stress tensor}
Considering the case with fixed R-R charge and keeping the transverse $S^5$ undeformed, then making a KK reduction of the black D3-brane along $S^5$ to obtain the 5-dimensional effective theory on the D3 brane world-volume. To analyze the hydrodynamics description, it's convenient to rewrite the world-volume metric $ds_5^2$ into the ingoing Eddington-Finkelstein coordinate
\be
    ds_5^2 = 2\Delta_-^{-\frac 3 4} dv\ dr - \Delta_+ \Delta_-^{-\frac 1 2} dv^2 + \Delta_-^{\frac 1 2} dy_i^2,
\ee
where $v$ is the ingoing coordinate
\be
    v = t+r_* \equiv t + \int \frac{dr}{\Delta_+ \Delta_-^{\frac 1 4}}.
\ee
Renaming $\{v, y_i\}\equiv x^\mu$ ($\mu=0,1,2,3$) and further performing a Lorentz boost transformation with constant 4-velocity $u_\mu$ results in a metric of the form:
\be
    ds_5^{2} = -2 \Delta_-^{-\frac 3 4} u_\mu dx^\mu dr - \Delta_+\Delta_-^{-\frac 1 2} u_\mu u_\nu dx^\mu dx^\nu + \Delta_-^{\frac 1 2} P_{\mu\nu} dx^\mu dx^\nu
\ee
where $P_{\mu\nu}=u_\mu u_\nu + \eta_{\mu\nu}$ is the projecting operator. In addition, the Dirichlet boundary condition is adopted so that the induced metric $ds_5^2$ is flat on a time-like cut-off surface $r=R$, namely, a rescaling transformation of the coordinates is made
\be
    ds_5^{2} = -2 \frac{\Delta_-^{-\frac 3 4}}{\Delta_{+R}^{\frac 1 2}\Delta_{-R}^{-\frac 1 4}} u_\mu dx^\mu dr - \frac{\Delta_+\Delta_-^{-\frac 1 2}}{\Delta_{+R}\Delta_{-R}^{-\frac 1 2}} u_\mu u_\nu dx^\mu dx^\nu + \frac{\Delta_-^{\frac 1 2}}{\Delta_{-R}^{\frac 1 2}} P_{\mu\nu} dx^\mu dx^\nu, \label{rescaledm}
\ee
where $\Delta_{\pm R} \equiv \Delta_\pm(R)$.

Further, from the effective theory on the world-sheet, the renormalized stress tensor on the cut-off surface $r=R$ can be obtained as~\cite{Emparan:2013ila}
\be\label{stresst1}
    T_{\mu\nu} = \frac{1}{\kappa_5^2} \left[ R^5 (K_{\mu\nu} - h_{\mu\nu}K) + (Q-5R^4) h_{\mu\nu} \right]
\ee
where $K_{\mu\nu}$ is the extrinsic curvature of the hypersurface $r=R$ and $h_{\mu\nu}$ the induced metric on $r=R$, which is flat $\eta_{\mu\nu}$ by construction. The stress tensor satisfies the conservation equation $\nabla^\mu T_{\mu\nu} = 0$ which is determined by the constraint equations of the Einstein equation on the world-volume, where $\nabla$ is the covariant derivative compatible with the reduced metric $h_{\mu\nu}$.
\omits{Here the greek indices $\mu$ and $\nu$ only cover the 3+1 dimensions, but the capital letter indices $A$ and $B$ are used to indicate all dimensions except for $r$, so that $ds^2=h_{AB}dX^A dX^B + \frac{dr^2}{\Delta_+\Delta_-}$, $h_{AB}\neq\eta_{AB}$, with $X^A\equiv\{t,x_i,\Omega_5\}$. The calculation of $T_{\mu\nu}$ requires firstly the calculation of the extrinsic curvature tensor $K_{AB}$, which is defined as the Lie derivative of $h_{AB}$ along the $r$ direction:
\be
K_{AB} &=& -\frac{1}{2} (\Lie_r h_{AB}) |_{r=R} \nno\\
K_{AB}dX^A dX^B &=& -{\frac 1 2} \sqrt{\Delta_{+R}\Delta_{-R}} \left[ -\left( \frac{\Delta_+^\prime(R)}{\Delta_{+R}} + \frac{(\Delta_-^{-1/2})^\prime(R)}{\Delta_{-R}^{-1/2}} \right) u_\mu u_\nu dx^\mu dx^\nu \right.\nno\\
&& \qquad\qquad\qquad\qquad \left. {} + \frac{(\Delta_-^{-1/2})^\prime(R)}{\Delta_{-R}^{-1/2}} P_{\mu\nu} dx^\mu dx^\nu + 2R \delta_5 d\Omega_5^2 \right]
\ee

Then by taking its trace in 4+5 dimensions:
\be
K &=& -\frac{1}{2} \sqrt{\Delta_{+R}\Delta_{-R}} \left( \frac{\Delta_{+R}^\prime}{\Delta_{+R}} + \frac{\Delta_{-R}^\prime}{\Delta_{-R}} + \frac{10}{R} \right),
\ee}
It is shown in that the explicit expression for $T_{\mu\nu}$ can also be written in a perfect fluid form
\be\label{stresst2}
    T_{\mu\nu} &=& \epsilon_c u_\mu u_\nu + p_c P_{\mu\nu}
\ee
with the fluid energy density $\epsilon_c$ and the fluid pressure $p_c$ being
\be
    \epsilon_c &=& \frac{R^4}{\kappa_5^2} \left( 5 -\frac{Q}{R^4} -5\sqrt{\Delta_{-R}\Delta_{+R}} -\frac{3}{4}R\sqrt{\frac{\Delta_{+R}}{\Delta_{-R}}}\Delta_{-R}^\prime \right), \nno\\
    p_c &=& -\epsilon_c + \frac{R^5}{\kappa_5^2} \frac{\Delta_{-R}\Delta_{+R}^\prime - \Delta_{+R}\Delta_{-R}^\prime}{2\sqrt{\Delta_{+R}\Delta_{-R}}}.
\ee
\omits{The second formula can be rewritten in terms of variables  $\vl$, $\de$ and $R_c$:
\be
    \epsilon_c + p_c = \frac{2\vl^4}{\kappa_5^2} \de R_c^{-\frac 1 2} (1-\de)^{-\frac 1 2} (\de+R_c-\de R_c)^{-\frac 1 2},
\ee
where $\vl^2 = r_+ r_-$ has been held fixed, $R_c = \Delta_{+R}$ is the parameter specifying the boundary's position, and the parameter $\delta_e = \Delta_-(r_+)$ characterize the extremality.}
It is straightforward to check that the Euler relation $\epsilon_c + p_c = T_c s_c$ is satisfied where the redshifted local temperature $T_c$ and the entropy density $s$ on the cut-off surface are respectively
\be\label{T_c&s_c}
    T_c &=& \frac{1}{\pi r_+} \Delta_-(r_+)^{\frac 1 4} \Delta_{+R}^{-\frac 1 2} \Delta_{-R}^{\frac 1 4},\\\nno
    s_c &=& \frac{2\pi}{\kappa_5^2} r_+^5 \Delta_-(r_+)^{\frac 3 4} \Delta_{-R}^{-\frac 3 4}.
\ee
Besides, the first law of thermodynamics $T_c\delta s_c=\delta\epsilon_c$ is held, where the variation is acting on $r_+$ and since the R-R charge $Q$ is fixed, so $\delta r_-=-r_-\delta r_+/r_+$.

\subsection{Hydrodynamic fluctuations}\label{subsec:hydro fluct}
Once having identified the leading-order effective hydrodynamic description to a perfect fluid living on the cut-off surface, one may now introduce hydrodynamic fluctuations by letting the 4-velocity $u_\mu$ vary slowly from point to point, i.e. $u(x^\mu) = u^{(0)}+u^{(1)}(x^\mu)+...$. The first order term $u_\mu^{(1)}$ will be finite but small with respect to the constant term $u_\mu^{(0)}$. Moreover, with $\vl^2=r_+r_-$ and $R$ both held fixed, only the horizon radius $r_+$ is allowed to fluctuate (just like the case of perturbing a neutral black hole), i.e. $r_+(x^\mu) = r_+^{(0)}+ r_+^{(1)}(x^\mu)+...$. Under such approximations, the reduced metric and the stress tensor can be written in the gradient expansion as
\be\label{1sthydro}
    ds_5^2 &=& ds_5^{(0)\ 2} + ds_5^{(1)\ 2} +... , \\
    T_{\mu\nu} &=& T^{(0)}_{\mu\nu} + T^{(1)}_{\mu\nu} + ...
\ee
where $ds_5^{(0)\ 2}$ previously appeared in Eq.~(\ref{rescaledm}) now becomes a seed metric for calculating higher order corrections of the metric and the perturbed $ds_5^2$ is required to obey the Einstein equation on the worldsheet.
\omits{\be
    ds_5^{(0)\ 2} = -2 \frac{\Delta_-^{-\frac 3 4}}{\Delta_{-R}^{-\frac 1 4}\Delta_{+R}^{\frac 1 2}} u_\mu dx^\mu dr - \frac{\Delta_+\Delta_-^{-\frac 1 2}}{\Delta_{+R}\Delta_{-R}^{-\frac 1 2}} u_\mu u_\nu dx^\mu dx^\nu + \frac{\Delta_-^{\frac 1 2}}{\Delta_{-R}^{\frac 1 2}} P_{\mu\nu} dx^\mu dx^\nu, \label{ds0}
\ee}
Thanks to the symmetry of the cutoff surface, the first order correction can be classified into the tensor mode $\tt$, the vector mode $\vv$ and scalar modes $\ss_1$, $\ss_2$, $\ss_3$ as
\be
ds_5^{(1)\ 2} &=& \tt \sigma_{\mu\nu} dx^\mu dx^\nu + \left[ \theta \left( \ss_1 u_\mu u_\nu + \frac{1}{3}\ss_2 P_{\mu\nu} \right) + \vv a_{(\mu}u_{\nu)} \right] dx^\mu dx^\nu + 2\ss_3 \theta u_\mu dx^\mu dr \label{ds1}
\ee
with the shear tensor $\sigma_{\mu\nu} = P_\mu{}^\alpha P_\nu{}^\beta (\nabla_{(\alpha} u_{\beta)} - \frac{1}{3}\theta P_{\alpha\beta} )$, the expansion $\theta = \nabla_\mu u^\mu$ and the acceleration $a^\mu = u^\nu \nabla_\nu u^\mu$ of the dual fluid. $\tt$, $\vv$, $\ss_1$, $\ss_2$ and $\ss_3$ are functions depending on $r$, $r_+$ (thus on $x^\mu$) and $R$. Their explicit expressions can be found in \cite{Emparan:2013ila}.

Similarly, the leading-order stress tensor $T^{(0)}_{\mu\nu} = \epsilon_c u_\mu u_\nu + p_c P_{\mu\nu}$ will no longer be solution to the conservation equation (see section below). Its first order correction term $T^{(1)}_{\mu\nu}$ is expected to be a viscous stress tensor for a neutral fluid, i.e.,
\be
    T^{(1)}_{\mu\nu} = -2\eta_c \sigma_{\mu\nu} -\zeta_c \theta P_{\mu\nu},
\ee
where $\eta_c$ is the shear viscosity and $\zeta_c$ is the bulk viscosity on the cutoff surface, they are
\be
\eta_c &=& \frac{1}{2\kappa_5^2} r_+^5 \frac{\Delta_-(r_+)^{\frac 3 4}}{\Delta_{-R}^{\frac 3 4}}, \\
\zeta_c &=& \frac{40}{3\kappa_5^2} \frac{R^8 r_+^{29}}{\mathfrak{s}^2} \Delta_{-R}^{\frac 5 4} \Delta_-(r_+)^{\frac {11} {4}}.
\ee
with $\mathfrak{s}\equiv \frac{\vl^{16}}{(1-R_c)(1-\de)^2} \left[ 3\de+6R_c+2\de^2-4\de R_c -2\de^2R_c \right]$, where $\vl^2 = r_+ r_-$ has been held fixed, $R_c = \Delta_{+R}$ is the parameter specifying the boundary's position, and the parameter $\delta_e = \Delta_-(r_+)$ characterize the extremality.
\omits{In the following, the notations $\epsilon_c$, $p_c$, $\eta_c$ and $\zeta_c$ will be used if it causes no confusion. The stress tensor of the dual fluid living on the $r=R$, up to the first order correction, is that of a neutral viscous fluid:
\be
T_{\mu\nu} = \epsilon_c u_\mu u_\nu + p_c P_{\mu\nu} - 2\eta_c \sigma_{\mu\nu} -\zeta_c \theta P_{\mu\nu}.
\ee
}

\section{The acoustic black hole }\label{sec:acoustic bh}

\subsection{Bernoulli equation and continuity equation}
In order to study the propagation of normal mode perturbations in the dual fluid, we will first separate the conservation equation $\nabla^\mu T_{\mu\nu} = 0$ into the longitudinal part and the transverse part. The longitudinal part gives the continuity equation, which will in turn simplify into an acoustic metric, while the transverse part gives the Bernoulli equation, which will be a gauge fixing condition.

Following similar steps described in \cite{Visser:2010xv} for relativistic perfect fluid, we start by supposing the 4-velocity field to be irrotational, then the fluid velocity can be expressed as
\be
u_\mu = \frac{\nabla_\mu \psi}{\sqrt{-\nabla_\nu\psi\nabla^\nu\psi}},
\ee
which automatically satisfies $u_\mu u^\mu = -1$ and $u\wedge du=0$, while the velocity potential $\psi$ is uniquely defined up to a pre-factor. Since the Dirichlet boundary condition has been adopted, the indices can be raised or lowered with $\eta_{\mu\nu}$ safely.

Further using $\nabla_\mu u_\nu = P^\alpha{}_\nu \frac{\nabla_\mu\nabla_\alpha\psi}{\norm{\nabla\psi}}$,  $a_\mu = u^\nu \nabla_\nu u_\mu = -P_\mu{}^\alpha \nabla_\alpha \log \norm{\nabla\psi}$, where $\norm{\nabla\psi} \equiv \sqrt{-\nabla_\nu\psi\nabla^\nu\psi}$, then the conservation equation projected along the transverse direction simplifies as (for convenience, we omit the subscript of $p_c$ and $\epsilon_c$ throughout the rest of the paper)
\be\label{transverseeq}
0 &=& P^{\nu\lambda} \nabla^\mu T_{\mu\nu} \nno\\
&=& (\epsilon+p) P^{\nu\lambda} u_\mu \nabla^\mu u_\nu + \nabla^\lambda p + ... \nno\\
&=& (\epsilon+p) \left[ P^{\nu\lambda} a_\nu + P^{\mu\lambda} \nabla_\mu \left( \int_0^p \frac{dp}{\epsilon(p)+p} \right) \right] + ... \nno\\
&=& (\epsilon+p) P^{\alpha\lambda} \nabla_\alpha \left[ -\log\norm{\nabla\psi} +\int_0^p \frac{dp}{\epsilon(p)+p} \right] + ...
\ee
Note that for any arbitrary scalar function $f$ depending only on $\psi$, an equality $P^{\alpha\lambda} \nabla_\alpha f(\psi)=0$ always holds. Hence Eq.~(\ref{transverseeq}) gives at leading order
\be
\log\norm{\nabla\psi} -\int_0^p \frac{dp}{\epsilon(p)+p} +f(\psi) =0
\ee

A change of variable $\psi \rightarrow \int_0^\psi \exp{-f(\psi^\prime)}d\psi^\prime$ can absorb the $f(\psi)$ term, so as to obtain the general relativistic Bernoulli equation
\be\label{bernoullieq}
\norm{\nabla\psi} = \exp\left(\int_0^p \frac{dp}{\epsilon(p)+p}\right)
\ee

Next, the longitudinal mode of the conservation equation
\be\label{longitudinaleq}
0 &=& u^\nu \nabla^\mu T_{\mu\nu} \nno\\
&=& -(\epsilon+p) \left[ \theta + \frac{u_\mu \nabla^\mu\epsilon}{\epsilon+p} \right] + ... \nno\\
&=& -(\epsilon+p) \left[ \theta + \left(\exp\left(-\int_0^\epsilon \frac{d\epsilon}{\epsilon+p(\epsilon)}\right)\right) u_\mu \nabla^\mu \left( \exp\int_0^\epsilon \frac{d\epsilon}{\epsilon+p(\epsilon)} \right) \right] + ...
\ee
Introducing a reduced ``particle number density'' of the fluid
\be\label{numberd}
n=\frac{N}{N(\epsilon=0)} \equiv \exp{\int_0^\epsilon \frac{d\epsilon}{\epsilon+p(\epsilon)}},
\ee
then Eq.~(\ref{longitudinaleq}) at leading order gives the continuity equation
\be\label{continuityeq}
\nabla_\mu(n u^\mu)=0
\ee

\subsection{Acoustic metric}
Let us now linearize the fluid equations by adding infinitesimal normal mode perturbations as
\be\label{normalmode}
\psi=\bar\psi+\delta\psi, \quad
p=\bar p+\delta p, \quad
\epsilon=\bar\epsilon+\delta\epsilon \quad {\rm and} \quad
n=\bar n+\delta n
\ee
Then Eq.~(\ref{bernoullieq}) and Eq.~(\ref{numberd}) are linearized as:
\be
\frac{\delta\norm{\nabla\psi}}{\norm{\nabla\bar\psi}} = \frac{\delta p}{\bar\epsilon+\bar p}, \qquad
\frac{\delta n}{\bar n} = \frac{\delta\epsilon}{\bar\epsilon+\bar p},
\ee
where $\delta p$ and $\delta\epsilon$ are related through the speed of sound via $c_s^2 = \frac{\delta p}{\delta\epsilon}$ and $c_s$ is~\cite{Emparan:2013ila}:
\be\label{c_s}
    c_s^2 = \frac{2\de^2+5\de R_c - 8\de^2R_c +2R_c^2 -8\de R_c^2 +6\de^2R_c^2}{R_c(3\de+2\de^2+6R_c-4\de R_c -2\de^2R_c)}
\ee
where $R_c = \Delta_{+R}$ is the parameter specifying the boundary's position, and $\delta_e = \Delta_-(r_+)$ describes the deviation from extremality. Thus $\delta n = \frac{\bar n}{c_s^2} \frac{\delta\norm{\nabla\psi}}{\norm{\nabla\psi}} = -\frac{\bar n}{c_s^2} \frac{\bar u^\nu\nabla_\nu\delta\psi}{\norm{\nabla\bar\psi}}$. Together with $\delta u^\mu = P^{\mu\nu}\frac{\nabla_\nu\delta\psi}{\norm{\nabla\bar\psi}}$, the continuity equation~(\ref{continuityeq}) is finally linearized as (in the following, the bars are also neglected for convenience)
\be\label{eomphonon}
    \nabla_\mu\left( \frac{n}{\norm{\nabla\psi}} \left( -\frac{1}{c_s^2}u^\mu u^\nu + P^{\mu\nu} \right) \nabla_\nu\delta\psi \right) = \mathcal{O}(\partial^3 \delta\psi)
\ee
which is the equation of motion of a phonon propagating in a background acoustic metric:
\be\label{acogeo}
    ds_{\rm{ac}}^2 = \frac{n^2}{(\epsilon+p)c_s} \left( -c_s^2 u_\mu u_\nu + P_{\mu\nu} \right) dx^\mu dx^\nu
\ee

In order to eliminate the off-diagonal terms, one defines a new time coordinate $d\tau = dx^0 + \frac{(1-c_s^2)u_0 u_i}{(1-c_s^2)u_0^2-1}dx^i$:
\be
    ds_{\rm{ac}}^2 = \frac{n^2}{(\epsilon+p)c_s}\left[ \left( (1-c_s^2)u_0^2-1 \right)d\tau^2 + \left( \delta_{ij} -\frac{(1-c_s^2)u_i u_j}{(1-c_s^2)u_0^2-1} \right) dx^i dx^j \right]
\ee

\section{On the bulk/boundary duality}\label{sec:bulk boundary}
\subsection{Connection between the dynamics}\label{subsec:dynamics}
Now we will analyze the dynamical connection between the above acoustic black hole and the bulk black D3-brane. Using the Codazzi equation
\be \label{codazzi}
\nabla^\alpha\left(K_{\alpha\beta}-h_{\alpha\beta}K\right)=R_{AB}n^B h^A_\beta,
\ee
where the bulk 5-dimensional Ricci tensor $R_{AB}=5\left(D_A\varphi D_B\varphi+\nabla_A\nabla_B\varphi\right)-Q^2 e^{-10\varphi}g_{AB}$ and $D_A$ is the covariant derivative compatible with the 5-dimensional bulk metric $g_{AB}$ and $n^B$ is the unit normal vector of the cutoff surface. In addition, the dilaton field $\varphi=\ln r$ and it is required to be a constant on the cutoff surface, i.e. $h^A_\mu D_A\varphi=0$, and hence $h^A_\mu D_A r=0$. Consequently, the right hand side of eq.(\ref{codazzi}) equals zero, which gives the conservation equation of the stress tensor at the cutoff surface $\nabla^\mu T_{\mu\nu} = 0$. This builds up the dynamical connection between the fluid and the bulk black D3-brane. Furthermore, the acoustic black hole is formed from the normal mode fluctuation eq.(\ref{normalmode}) of the fluid, then dynamical connection between the acoustic black hole and the bulk gravity is
\be \label{dynamics}
\nabla^\mu\delta T_{\mu\nu}=\frac{R^5}{\kappa_5^2}\delta R_{AB}n^B h^A_\nu,
\ee
where the equations on the left hand side govern the dynamics of the phonon eq.(\ref{eomphonon}) and hence determine the acoustic geometry eq.(\ref{acogeo}). While the equations on the right hand side describe the quasinormal excitations in the bulk gravity.

\subsection{Temperature of sound horizon vs. temperature of real black hole}\label{subsec:Hawking T}
Without loss of generality, we can choose the coordinates to make the fluid flowing along the $x^3 \equiv z-$direction, namely,
\be
u_\mu = (u_0, 0, 0, u_z),
\ee
then the acoustic metric is simplified into
\be
    ds_\text{ac}^2 = \frac{n^2}{(\epsilon_c+p_c)c_s} \left( -\left(1-(1-c_s^2)u_0^2 \right) d\tau^2 + dx^a dx_a + \frac{c_s^2}{1-(1-c_s^2)u_0^2}dz^2 \right), \label{acousticm}
\ee
where index $a$ and $b$ run in $x^1$ and $x^2$, and there is a $SO(2)$ symmetry on the $x^1-x^2$ plane. A coordinate singularity will appear where the velocity $u_\mu$ grows slowly and reaches a critical value. As we already stated in section~\ref{subsec:hydro fluct}, the non-uniform behaviour of $u_\mu$ modifies the bulk metric via back-reaction, creating the higher order dissipative terms in the stress tensor. Infinitesimal perturbations made to those terms would produce the quasi-normal modes that will be discussed in the next subsection. For instance, we will focus on thermodynamic properties as seen by the sound normal modes, i.e. by phonons on the cut-off surface.

The horizon temperature near the singularity $u_0^2 = \frac{1}{1-c_s^2}$ i.e. $u_z^2= \frac{c_s^2}{1-c_s^2}$ is calculated:
\be\label{T_sh}
    T_\text{sh} = \frac{\sqrt{1-c_s^2}}{2\pi}\vert \partial_z u_z \vert
\ee
But this sound horizon is not always present, it requires $1-c_s^2>0$. By examining the expression of the speed of sound $c_s$ (eq.(\ref{c_s})), we noticed the following:
\begin{itemize}
    \item In the limit $R \rightarrow \infty$, i.e. $R_c \rightarrow 1$, $c_s^2$ simplifies to $\frac{2-3\de}{6-\de}$, which monotonically decreases from $\frac 1 3$ at $\de=0$ to $-\frac 1 5$ at $\de=1$. A Gregory-Laflamme instability will develop when $c_s^2<0$, i.e. $\de > \frac 2 3$ (charge sufficiently small), whereas $1-c_s^2$ will always be positive, indicating the formation of a sound horizon near $u_z = \frac{c_s}{\sqrt{1-c_s^2}}$.
    \item On the contrary, in the limit $R \rightarrow r_+$, i.e. $R_c \rightarrow 0$, the cut-off surface approaches the outer horizon. If the black brane is non-extremal, i.e. $\de \neq 0$, then $c_s^2$ diverges, assuring the stability of the fluid, while forbidding the formation of a sound horizon. The reason why the system becomes non-relativistic may be as the following. By imposing Dirichlet boundary condition at $r=R$, there'll be less `room' for fluctuations between the cut-off surface and the horizon, thus it `rigidifies' the system~\cite{Emparan:2013ila}.
    \item The last case is to take both $\de$ and $R_c$ to $0$, while the ratio $\frac{\de}{R_c}$ may not tend to $0$, then $c_s^2$ actually goes as $\frac{2(\de/R_c)^2 +5\de/R_c +2}{3\de/R_c+6}$. Again, the fluid will always be stable, whereas the condition of sound horizon formation $c_s^2<1$ turns out to be $\frac{\de}{R_c} < 1$. In the case of near extremality $\de \rightarrow 0$, the speed of sound $c_s^2 \rightarrow \frac 1 3$ when $\frac{\de}{R_c} \rightarrow 0$.
\end{itemize}

One of the novel and key findings in~\cite{Ge:2015uaa} was that the Hawking-like temperature of the acoustic black hole is really connected to the Hawking temperature of a real black hole. Next, we will analyze the relation between the Hawking temperature $T_c$ and the temperature $T_\text{sh}$ of the sound horizon via the term $\partial_z \ln r_+$.

In order to relate the velocity gradient in $T_\text{sh} = \frac{\sqrt{1-c_s^2}}{2\pi}\vert \partial_z u_z \vert$ to the radius gradient $\partial_z r_+$, we'll use the relation $\nabla_\mu\ln s_c = \theta u_\mu - c_s^{-2}a_\mu$. In the present case, $\theta = \partial_z u_z$, $a_z = u_z\partial_zu_z$, so that $\partial_z\ln s_c = (1-c_s^{-2})u_z\partial_zu_z$, where $u_z\vert_{z_{sh}} = \frac{c_s}{\sqrt{1-c_s^2}}$. Check Eq.(\ref{T_c&s_c}) for the explicit value of the entropy density $s_c$ and the Hawking temperature $T_c$, then we have:
\be
T_\text{sh} &=& \frac{c_s}{2\pi} \partial_z\ln s_c \nno\\
            &=& \frac{c_s}{2\pi} \left( 2+\frac{6}{\de}-\frac{3}{\Delta_{-R}} \right) \partial_z \ln r_+ \\
\partial_z \ln T_c
    &=& \left( \frac{2}{\de}+\frac{2}{\Delta_{+R}}+\frac{1}{\Delta_{-R}} -6 \right) \partial_z \ln r_+.
\ee

First, we step back a little and notice immediately that when the cut-off surface $r=R$ recedes to infinity, $c_s^2 \rightarrow \frac{2-3\de}{6-\de}$, an equality holds at that limit:
\be
\lim_{R \rightarrow \infty} 2\pi c_s T_\text{sh} = \frac{2-3\de}{\de} \partial_z \ln r_+ = \lim_{R \rightarrow \infty} \partial_z \ln T_c\vert_{z_{sh}}
\ee
Later, a complete calculation shows that the quotient of $\frac{2}{\de}+\frac{2}{\Delta_{+R}}+\frac{1}{\Delta_{-R}} -6$ by $2+\frac{6}{\de}-\frac{3}{\Delta_{-R}}$ equals exactly to the expression of $c_s^2$, thus giving the universal equation
\be\label{TT}
    2\pi c_s T_\text{sh} = \partial_z \ln T_c \vert_{z_\text{sh}}
\ee
which is independent of the position of the cut-off surface $R$ and the extremality parameter $\delta_e$ (thus the R-R charge). This is different from the situation in the asymptotically AdS spacetime in which the eq.(\ref{TT}) only holds at the AdS boundary.

\subsection{Sound mode vs. scalar quasi-normal mode}
Recall that the hydrodynamic fluctuations $u_\mu^{(1)}$ on the boundary create a back-reaction $ds_5^{(1)}$ in the bulk Eq.~(\ref{ds1}). Further performing the quasinormal mode perturbation to the hydrodynamically perturbed black D3-brane geometry eq.(\ref{1sthydro}), one obtains the bulk quasinormal modes. It was shown in~\cite{Ge:2015uaa} that there is a phonon/scalar quasinormal mode duality. In the present case, we have
\be
    \delta g_{00}^{(1)} &=& \left[ \tt'\left( u_0a_0 -\frac{1}{3}\theta P_{00} \right) + \theta\left( \ss_1'u_0^2 + \frac{1}{3}\ss_2'P_{00} \right) + \vv'u_0 a_0 \right] \delta r_+ \nno\\
    \delta g_{zz}^{(1)} &=& \left[ \tt'\left( u_za_z + \partial_zu_z -\frac{1}{3}\theta P_{zz} \right) +\theta\left( \ss_1'u_z^2 +\frac{1}{3}\ss_2'P_{zz} \right) + \vv'a_zu_z \right] \delta r_+ \nno\\
    \delta g_{0z}^{(1)} &=& \left[ \tt'\left( \frac{1}{2}(u_0a_z + u_za_0 + \partial_zu_0) -\frac{1}{3}\theta P_{0z} \right) + \theta\left( \ss_1'u_0u_z +\frac{1}{3}\ss_2'P_{0z} \right) + \vv'\frac{1}{2} (u_0a_z + u_za_0) \right] \delta r_+ \nno\\
    \delta g_{ab}^{(1)} &=& \frac{1}{3}(\ss_2'- \tt')\theta\eta_{ab} \delta r_+, \quad {\rm when} \quad a,b \neq 0,z,
\ee
where the variation $\delta$ is acting on $r_+$. They are the only non-zero components of $\delta g_{\mu\nu}^{(1)}$ under the condition $u=(u_0, 0, 0, u_z)$ and can be assembled into a gauge invariant scalar field $Z_c$ (which is the scalar channel of the quasinormal mode) thanks to the $SO(2)$ symmetry in the background geometry
\be
Z_c &=& u_z^2 \delta g_{00}^{(1)} + u_0^2 \delta g_{zz}^{(1)} -2u_0u_z \delta g_{0z}^{(1)} +\delta g_{11}^{(1)} + \delta g_{22}^{(1)} \\\nno
    &=& \theta \ss_2'\delta r_+
\ee
Then by using the entropy relation $\nabla_\mu\ln s_c = \theta u_\mu - c_s^{-2}a_\mu$, we can also find the connection between the sound mode $Z_c$ and the phonon $\delta\psi$ propagating in the acoustic black hole via the following equation
\be
\left( (\frac{6}{\de} -\frac{3}{\Delta_{-R}} +2)(\partial_\mu \ln r_+) + \partial_\mu \right) \frac{P^{\mu\nu}\partial_\nu \delta\psi}{\norm{\nabla\psi}} = - (\frac{6}{\de} -\frac{3}{\Delta_{-R}} +2) u^\mu\partial_\mu \left( \frac{Z_c}{\theta r_+\ss_2'} \right),
\ee
which is the phonon/scalar quasinormal mode duality in the black D3-brane.

\section{Conclusions}\label{sec:conclusion}
In this paper, we studied a new analogous gravitational system, namely, the acoustic black hole appeared in the nonextremal black D3-brane in type IIB string theory. Although the nonextremal black D3-brane is in asymptotically flat spacetime, it also can be described as a hydrodynamic system in the low frequency and long wavelength limits, like the fluid/gravity duality in the AAdS. Using the methods in~\cite{Ge:2015uaa}, we showed that the acoustic black hole emerged on the timelike cutoff surface in this system also has a gravity dual, which is a hydrodynamically perturbed black D3-brane in the bulk. Based on the connection between the dynamics of the acoustic black hole and the bulk gravity, we found a universal relation between the Hawking-like temperature and the Hawking temperature. Besides, we found the field/operator duality relation, i.e., the phonon/scalar channel quasinormal mode duality still held in this flat spacetime case. Our results not only extended the applicability of the methods in~\cite{Ge:2015uaa} into a flat spacetime holography, but also revealed new information about the black D3-brane in the hydrodynamic limit, namely, the acoustic black hole can be used to describe the quasinormal modes fluctuations of the hydrodynamically perturbed black D3-brane. It would be interesting to study the entanglement entropy of the acoustic black hole and find its gravity dual description in the bulk black D3-brane, which may give more information about their dynamic connections.

\section*{Acknowledgement}
This work was supported by the National Natural Science Foundation of China (No.~11675272), the Open Project Program of State Key Laboratory of Theoretical Physics, Institute of Theoretical Physics, Chinese Academy of Sciences, China (No.~Y5KF161CJ1) and the Fundamental Research Funds for the Central Universities.





\end{document}